\documentclass[preprint,aps,superscriptaddress]{revtex4}
\usepackage{graphicx}
\usepackage{dcolumn}
\usepackage{bm}
\usepackage{times}

\begin{document}

\title{Experimental observation of superluminal group velocities in bulk
two-dimensional photonic bandgap crystals}

\author{D. R. Solli}
\affiliation{Department of Physics, University of California, Berkeley, CA 94720-7300.}
\author{C. F. McCormick}
\affiliation{Department of Physics, University of California, Berkeley, CA 94720-7300.}
\author{R. Y. Chiao}
\affiliation{Department of Physics, University of California, Berkeley, CA 94720-7300.}
\author{J. M. Hickmann}
\email{hickmann@loqnl.ufal.br}
\affiliation{Department of Physics, University of California, Berkeley, CA 94720-7300.}
\affiliation{Departamento de F\'{\i}sica, Universidade Federal de Alagoas, CidadeUniversit%
\'{a}ria, 57072-970, Macei\'{o}, AL, Brazil.}

\begin{abstract}
We have experimentally observed superluminal and infinite group
velocities in bulk hexagonal two-dimensional photonic bandgap crystals with
bandgaps in the microwave region. The group velocities depend on the
polarization of the incident radiation and the air-filling fraction of the
crystal.
\end{abstract}

\maketitle

\section{Introduction}

The superluminal propagation of wave packets with faster-than-$c$, infinite
and negative group velocities has been observed in a wide range of physical
systems, including both passive and active optical media \cite%
{Chiao1997,Nimtz1997,Boyd2002,Rochester,Wang2001}. Garrett and McCumber \cite%
{Garrett1970} first predicted that a Gaussian wave packet can propagate with
negligible distortion through a linear optical medium with a superluminal
group velocity, if its bandwidth is restricted to a narrow spectral region
of anomalous dispersion within an absorption or gain line. This prediction
was later experimentally verified for the propagation of picosecond laser
pulses through a GaP:N sample \cite{Chu1982}. Such superluminal behavior
does not violate relativistic causality. Any analytic function, such as a
Gaussian wave packet, contains sufficient information in its early tail to
reconstruct the entire wave packet with a superluminal pulse advancement,
but little distortion. Relativistic causality forbids only the \emph{front}
velocity, i.e., the velocity of any \emph{discontinuity}, from exceeding the
vacuum speed of light, $c$. It does not forbid the \emph{group} velocity of
a wavepacket from being superluminal.

The superluminal behavior of electromagnetic pulses can occur not only in
passive, \emph{absorptive} optical media, but also in passive, \emph{%
dissipationless} optical media, such as periodic dielectric photonic bandgap
(PBG) structures \cite{Yablonovitch}. Optical pulses whose bandwidths are
restricted to the PBG are exponentially damped not due to absorption, but
interference between multiple Bragg reflections from planes within the
transparent periodic dielectric structure. The superluminal propagation of
wavepackets through 1D periodic dieletric structures has been observed in
the optical spectral region \cite{Steinberg}. Similarly, superluminal
propagation has been observed for classical microwave pulses in a 1D
periodic dielectric stack placed inside a microwave waveguide \cite%
{Nimtz1994}. Other work has demonstrated superluminal behavior in the
frequency range of tens of MHz in a 1D photonic structure made of a sequence
of coaxial cables with alternating impedance values \cite{Hache2002}. No
reported experiment has yet demonstrated superluminal propagation of
electromagnetic wavepackets in 2D periodic structures, although we have
recently demonstrated for the first time clear experimental evidence for the
existence of exponential wave decay in such structures at microwave
frequencies \cite{Hickmann2002}. Here we report on experimental results
which imply that superluminal group velocities do occur in such structures.

Microwave measurements have certain advantages which we exploit here. First,
one can perform direct, precise phase measurements using a microwave vector
network analyzer (VNA). Second, the wavelength is sufficiently large so that
high-precision, well-fabricated 2D photonic crystals can be relatively
easily constructed. Third, one can more easily tailor the dispersive
properties of the PBG by changing structural parameters, such as the
structure air-filling fraction (AFF).

\section{Experiment}

The fabrication of our photonic crystals has been described in detail
elsewhere \cite{Hickmann2002}. In summary, we stack acrylic rods in a
hexagonal lattice and glue them in place with an acrylic solvent cement. We
constructed and tested crystals with as many as eighteen layers of rods, and
AFFs of 0.60, using rods with a 1/2-inch outer diameter. The crystals are
\textquotedblleft bulk\textquotedblright\ samples rather than
\textquotedblleft slabs\textquotedblright\ in the sense that their extruded
(non-periodic) dimension is many times the lattice spacing in length.

\begin{figure}\includegraphics{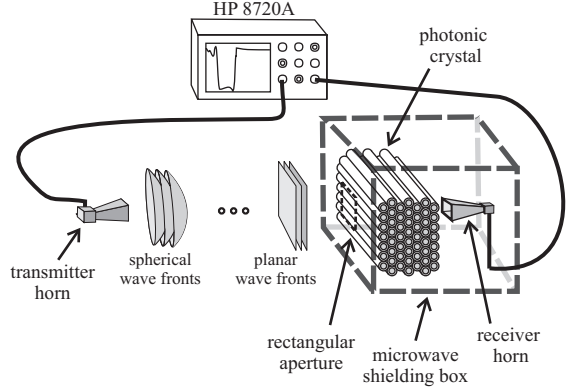}\caption{Experimental setup for microwave transmission and phase measurements.}\label{setup}\end{figure}\begin{figure}\includegraphics{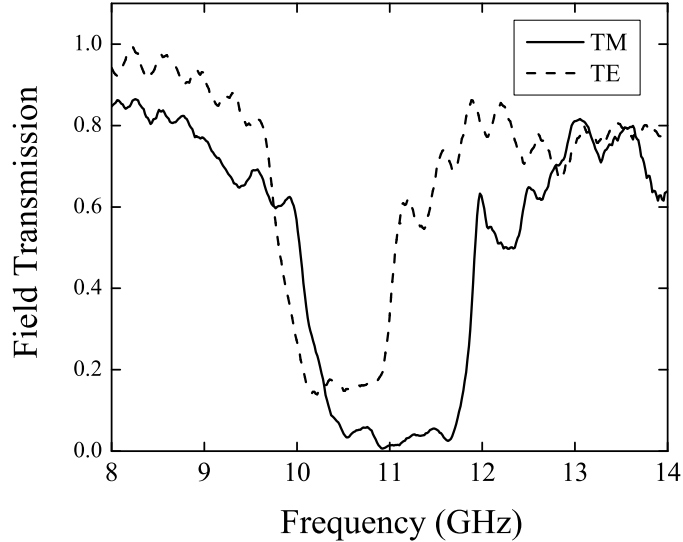}\caption{Transmission through an AFF equal to 0.60 crystal with 18 layers.  (a) TM polarization; (b) TE polarization.}\label{bandgap}\end{figure}\begin{figure}\includegraphics{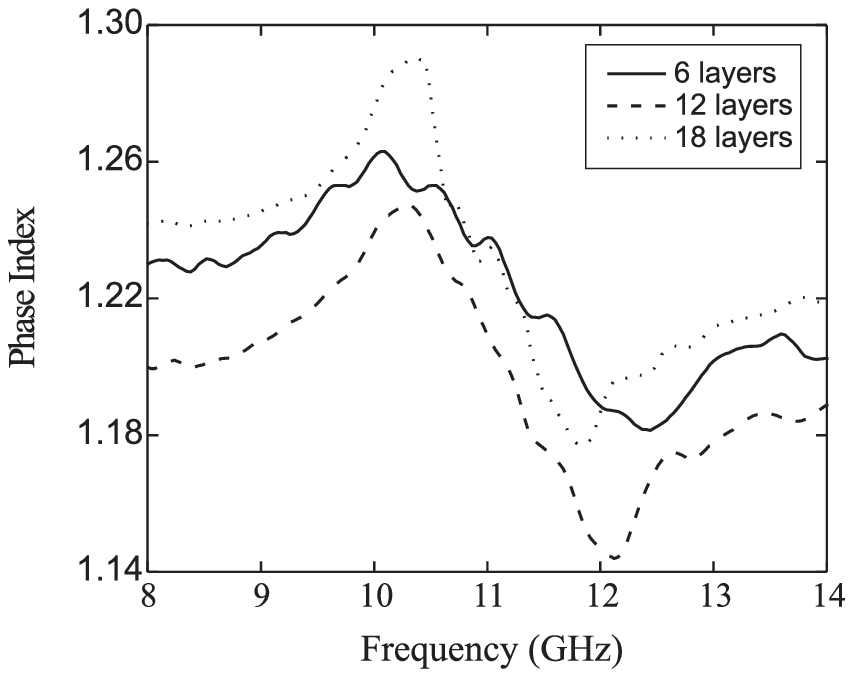}
\includegraphics{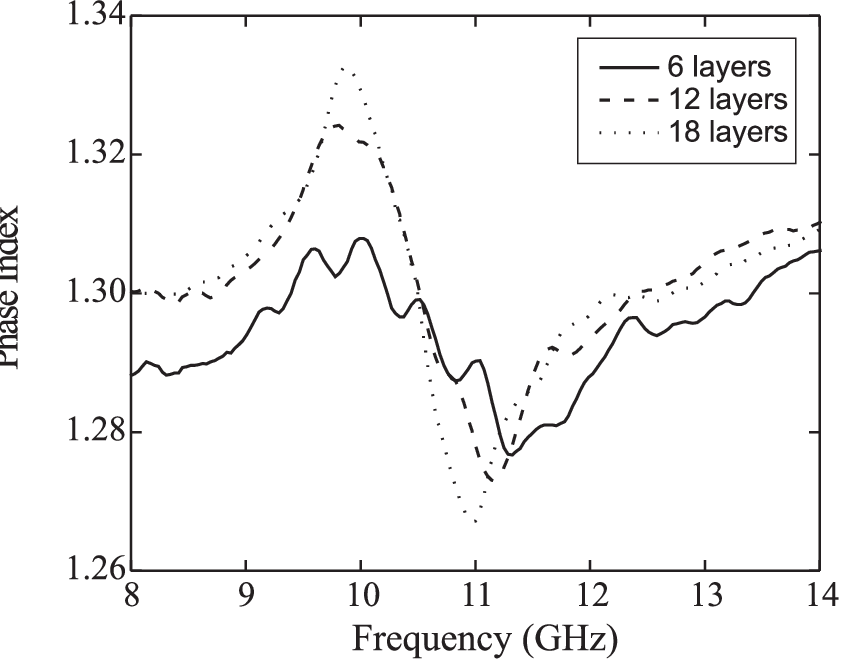}\caption{Phase index of refraction $n(\omega)$ for an AFF equal to 0.60 crystal with various layer numbers.  (a) TM polarization; (b) TE polarization.}\label{phaseindex}\end{figure}\begin{figure}\includegraphics{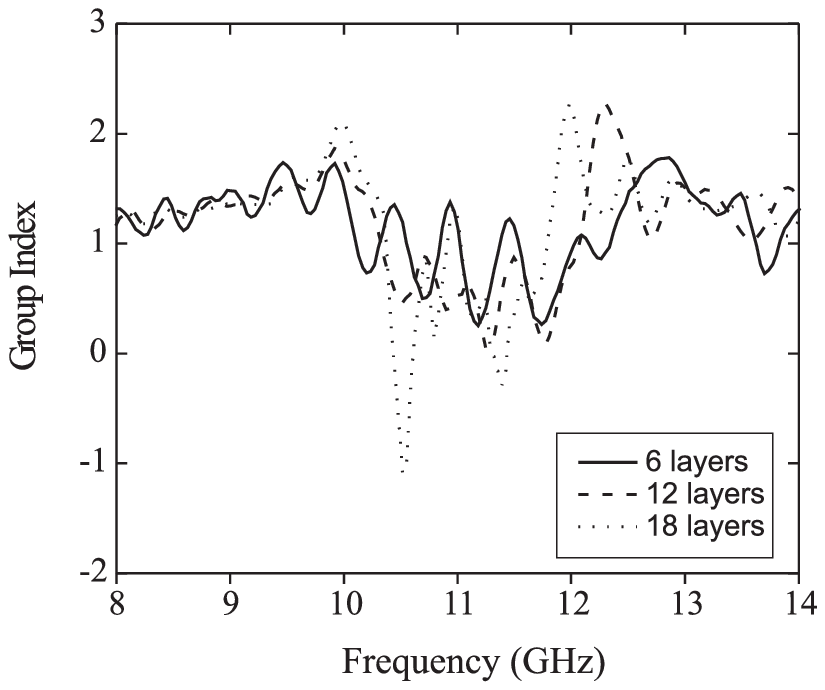}
\includegraphics{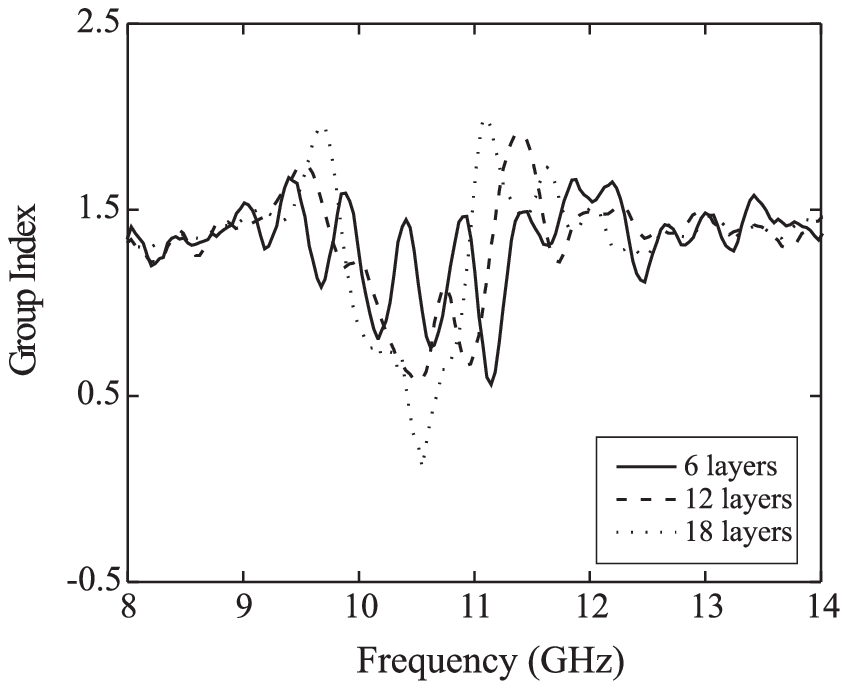}\caption{Group index of refraction $n_{g}(\omega)$ for an AFF equal to 0.60 crystal with various layer numbers.  (a) TM polarization; (b) TE polarization.}\label{groupindex}\end{figure}

In Fig. \ref{setup} we show a schematic of the experimental apparatus used
for the microwave transmission measurements. We used an HP 8720A VNA, which
is capable of measuring both the transmission and phase delay of microwaves
detected by a polarization-sensitive receiver antenna (horn). The VNA was
swept from 8 to 14 \thinspace GHz in 15 \thinspace MHz increments and was
connected to identical transmitter and receiver horns. The crystal and
receiver horn were placed 1.6 \thinspace m away in a box measuring 60
\thinspace cm on a side, with microwave-absorbing walls. Microwaves entered
the box through a 14 \thinspace cm $\times $ 17 \thinspace cm rectangular
aperture. These dimensions were chosen to minimize diffraction effects
through the aperture without allowing leakage around the crystal. With the
crystal removed and the rectangular aperture closed, the microwave signal at
the receiver was suppressed by more than 45 dB, indicating good shielding by
the box. As previously reported, our bulk crystal with AFF equal to 0.60
displays a strong bandgap at $\approx $ 11 \thinspace GHz \cite{Hickmann2002}%
.

To establish the reliability of the phase delay measurements, we placed
sheets of ordinary acrylic of varying widths in front of the receiver horn.
The phase delay relative to that of free space propagation was linear in the
transmitted microwave frequency, with a slope that depended on the thickness
of the acrylic. Using a simple model of phase delay given by

\begin{equation}
\Delta \phi (\omega )=\frac{\omega }{c}d(n(\omega )-1),
\label{phasedelay}
\end{equation}

\noindent where $d$ is the sheet thickness, we were able to extract a value
of $n=1.58$ for the index of refraction of acrylic, in good agreement with
the accepted value of $1.61$ for this frequency range \cite{Gray1972}.

The group velocity measurements were carried out by measuring, for both
polarizations, the spectral dependence of the phase delay for crystals with
different numbers of layers. The group index was calculated according to the
relation \cite{Garrett1970}

\begin{equation}
n_{g}(\omega _{0})=n(\omega _{0})+\omega _{0}\frac{\partial n}{\partial
\omega }|_{\omega _{0}},  \label{groupindexeqn}
\end{equation}

\noindent where $n(\omega)$ is the normal (phase) index of refraction and
the group index $n_{g}$ is defined as the ratio of the speed of light to the
group velocity. Thus, superluminal group velocities exist in any spectral
region where the group index is less than unity.

In order to calculate the group index, it was necessary to unwrap the phase
data obtained from the VNA, since it is sensitive to phase modulo $2\pi $.
We were able to determine when such phase slips occurred by observing that
the unwrapped phase delay at fixed frequencies was a monotonic function of
the number of layers except for sharp changes (phase slips) at certain
numbers. We adjusted the data by adding $2\pi \times m$ to each phase delay
spectrum, where $m$ is the cumulative number of sharp drops observed over
smaller layer numbers. We also confirmed the validity of this phase
adjustment by considering $\partial n/\partial \omega $, which should go to
zero far from the bandgap. We found that it did so if and only if we applied
this phase delay adjustment as described.

\section{Results and Discussion}

In Fig. \ref{bandgap}, we show the measured transmission bandgap of a
crystal with eighteen layers and AFF equal to 0.60, for both incident
microwave polarizations. We consider the plane of incidence to be the plane
of periodicity (perpendicular to the rods) and define the TE (TM)
polarization to have its electric field perpendicular (parallel) to this
plane. We normalized transmission measurements by removing the crystal and
measuring the total microwave signal at the receiving horn. We find that
this crystal exhibits anomalous dispersion within the bandgap, passing
briefly through zero dispersion at the bandgap edges, for both TE and TM
polarizations (see Fig. \ref{phaseindex}). We note that the region of
anomalous dispersion becomes increasingly well-defined as the number of
layers is increased.

From the phase delay data in Fig. \ref{phaseindex} we calculate the group
index using Eq. \ref{groupindexeqn}. The results are displayed in Fig. \ref%
{groupindex}. The region of superluminal $n_{g}$ extends over a wider
spectral band for the TM polarization than the TE polarization, and deepens
more rapidly as the number of layers is increased. This result is not
surprising given that the TM gap is known to form more rapidly than the TE
gap as the number of crystal layers is increased \cite{Hickmann2002}.

We did similar measurements (not shown here) using other crystals with AFF
equal to 0.32 and obtained qualitatively similar bandgaps, although they are
narrower and shallower for both polarizations, as expected \cite{Broeng1998}%
, and have slightly different center frequencies. Both crystal sets display
similar superluminal spectral regions, although we found that those with
higher AFF typically have greater bandwidth and yield superluminal group
velocities which are more superluminal.

\section{Conclusion}

We have experimentally demonstrated that superluminal including infinite 
group velocities can exist for analytic signals whose
spectral bandwidth lies within the bandgap of a two-dimensional hexagonal
photonic crystal. Superluminal phenomena are exhibited by these crystals for
both tested AFFs, although the effects are more significant (i.e., group
velocities become more superluminal and exist over a wider spectral range) for
structures of higher AFF.

This work was supported by ARO grant number DAAD19-02-1-0276. We thank the
UC Berkeley Astronomy Department, in particular Dr. R. Plambeck, for lending
us the VNA. JMH thanks the support from Instituto do Mil\^{e}nio de Informa%
\c{c}\~{a}o Qu\^{a}ntica, CAPES, CNPq, FAPEAL, PRONEX-NEON, ANP-CTPETRO.

\end{document}